# Vortex-Ring-Induced Internal Mixing Upon the Coalescence of Initially Stationary Droplets

X. Xia,[1,a)] C. He,[1,a)] D. Yu,[1] J. Zhao,[1] and P. Zhang[1,b)]

[1]*Department of Mechanical Engineering, The Hong Kong Polytechnic University, Hong Kong, P. R. China*

This study employs an improved volume of fluid method and adaptive mesh refinement algorithm to numerically investigate the internal jet-like mixing upon the coalescence of two initially stationary droplets of unequal sizes. The emergence of the internal jet is attributed to the formation of a main vortex ring, as the jet-like structure shows a strong correlation with the main vortex ring inside the merged droplet. By tracking the evolution of the main vortex ring together with its circulation, we identified two mechanisms that are essential to the internal-jet formation: the vortex-ring growth and the vortex-ring detachment. Recognizing that the manifestation of the vortex-ring-induced jet physically relies on the competition between the convection and viscous dissipation of the vortex ring, we further developed and substantiated a vortex-ring-based Reynolds number ($Re_J = \Gamma_{VD}/4\pi\nu$) criterion, where $\Gamma_{VD}$ is the circulation of the detached main vortex ring and $\nu$ is the liquid kinematic viscosity, to interpret the occurrence of the internal jet at various Ohnesorge numbers and size ratios. For the merged droplet with apparent jet formation, the average mixing rate after jet formation increases monotonically with $Re_J$, which therefore serves as an approximate measure of the jet strength. In this respect, stronger internal jet is responsible for enhanced mixing of the merged droplet.

---

a) The first and second authors contributed equally to this work.

b) Author to whom correspondence should be addressed. Electronic mail: pengzhang.zhang@polyu.edu.hk.

## I. INTRODUCTION

Droplet coalescence is a frequent event in many natural and industrial processes involving dispersed two-phase system of either gas–liquid or liquid–liquid[1-4]. A prominent example is the dense spray of liquid fuel in the combustor of a diesel[5,6], gas turbine[7] or rocket engine[8-10]. The large number density of droplets in the spray along with the flow non-uniformity in the combustor imply frequent droplet collisions. Among various outcomes of droplet collision, droplet coalescence is of paramount importance because it affects the number density and size distribution of colliding droplets, and in turn the subsequent spray and combustion characteristics[11]. Droplet coalescence is also an arguably important mechanism for explaining the size distribution of droplets in rain clouds[12,13].

Earlier studies on droplet coalescence were focused on quantifying its dependence on various collision parameters such as the Weber number, $We$, which measures the relative importance of droplet inertia compared to the surface tension, and the impact parameter, $B$, which measures the deviation of droplet trajectory from the head-on collision ($B = 0$). Specifically, the head-on collision of identical water droplets in atmospheric air results in either permanent coalescence at small $We$s or separation after temporary coalescence at large $We$s[14-17]. For the head-on collision of identical droplets of liquid hydrocarbons in atmospheric air, bouncing emerges at intermediate $We$s and coalescence occur either at smaller $We$ with minor droplet deformation or at larger $We$ with substantial droplet deformation. Furthermore, droplet coalescence can be significantly influenced by the ambient gas pressure as such coalescence is promoted at reduced gas pressures owing to the decreased inertia of the gas film separating the droplet surfaces.[11,18] By the same token, bouncing can be absent for hydrocarbon droplets when the gas pressure is sufficiently low[16,19,20].

Compared with the coalescence of two identical droplets, the coalescence of two droplets of unequal sizes is more practically relevant and has gained increasing attention in recent years. Ashgriz and Poo[14] experimentally found that water droplets tend to coalesce as their size difference increases. This was subsequently verified by Rabe et al.[17] and Testik[21] for water droplets and by Estrade et al.[22] for ethanol droplets. A recent study by Tang et al.[23] shows that the promotion of droplet coalescence by increasing the size difference is because of the increased viscous dissipation, and that the promotion exists not only for water droplets but also for hydrocarbon droplets.

The evolution of merged droplet interface is probably the most important aspect of droplet coalescence, and has been the major focus in the majority of the previous experimental or numerical studies[24-26]. Another important aspect is the internal mixing within the coalesced droplet, which is crucial to some microfluidics and propulsion systems involving biological or chemical reactions[27,28]. Because the internal mixing after coalescence is minimal for two identical droplets due to the intrinsic symmetry, it can be enhanced only by breaking the symmetry. Blanchette[29] numerically investigated the internal mixing within the coalesced droplet with varying surface tension and found that the surface tension variation results in faster mixing than the geometric effect. Anilkumar *et al.*[30] experimentally observed a jet-like mixing after the coalescence of two initially stationary water–glycerin solution droplets of unequal sizes. For the highly viscous droplets, noticeable mixing was however not observed in their experiments, and instead, the smaller droplet simply lodges onto the larger droplet after coalescence. Recently, Tang *et al.*[10] experimentally investigated the coalescence of two unequal-sized droplets with nonzero Weber numbers and observed the non-monotonic emergence of the jet-like mixing with increasing $We$. Specifically, the jet-like mixing emerged at small $We$s, in accordance with previous observations for initially stationary droplet coalescence [30, 31], but such a mixing was absent at higher $We$s as the increased droplet deformation absorbed the impact energy and therefore suppressed the jet formation. As $We$ further increases, the jet-like mixing reemerges while it shows a "two-leg" configuration.

Efforts have been made to understand the internal mixing subsequent to droplet coalescence. By using the front-tracking method, Liu *et al.*[32] simulated the mixing of unequal-size droplets with emphasis on elucidating the important role of the surface energy of the merged interface in forming the jet-like mixing. They found that increasing the droplet viscosity suppresses the mixing and increasing the size disparity promotes the mixing. The internal mixing was also reported in several other numerical studies employing the volume of fluid (VOF) method[33-35], the front-tracking method[31, 32] and the lattice Boltzmann method[36-38]. Nevertheless, detailed investigation and physical interpretation of the mixing process have not been adequately addressed in these studies.

Extensive studies have been done on the coalescence of a droplet into a liquid pool[39], where a vortex ring is formed below a critical $We$ that slightly depends on other factors for example the Froude number. A mechanism based on the generation of vorticity by accelerated flows at curved liquid free surfaces was proposed to explain the formation of the vortex ring[39-41]. According to the Helmholtz vorticity equation of a homogeneous fluid, vorticity cannot be generated in the interior of a fluid where the density gradient is absent. Consequently, vorticity must be generated at the gas–liquid interface where the Helmholtz equation however does not apply. On the gas–liquid interface, which can be approximated by a free surface of a viscous fluid, the vanishing shear stress requires a nonzero jump in velocity gradients and therefore generates a finite vorticity[39]. The vortex ring formation after coalescence is a synergic consequence of the vorticity generation on the gas-liquid interface, the vorticity transport (including accumulation and advection) to the droplet interior, and the vorticity dissipation within the droplet. To the knowledge of the authors, no attempts have been made to describe such a vorticity evolution within the coalesced droplet.

In the present study, we aim to numerically investigate the coalescence of initially stationary droplets of unequal sizes. The first focus of the study is the jet-like internal mixing upon the droplet coalescence and its parametric dependence on liquid viscosity, surface tension and size differential. The second is to analyze the evolution of concomitant vorticity and its interrelation with the internal mixing. Regardless of the fact that droplet collision with increasing $We$ may enhance the internal mixing, only initially stationary droplets (i.e. with a zero Weber number) was considered in the study based on the following considerations. First, the coalescence of initially stationary droplets can be treated as a leading order approximation to that with small $We$s. More importantly, the coalescence of initially stationary droplets occurs at the contact point of two spherical droplets, which do not deform prior to coalescence [42]. At non-zero $We$s, the complicated phenomena of draining the intervening gas film out of the gap between two colliding droplets [42] and the succeeding interface merging at the length scales of the van der Waals force pose great challenges to numerical simulation. The additional difficulty of dealing with droplet coalescence at non-zero Weber numbers will be avoided in the present study although it merits future studies.

The structure of the paper is organized as follows: the description of the problem, the specifications of numerical methods, and the analysis of grid independence are presented in Section II, followed by the comparison with the available experimental results in the literature, in Section III. The phenomenological description of the jet-like mixing for three representative liquids such as water, n-decane and n-tetradecane are discussed in Section IV. An analysis of vorticity evolution is presented in Section V to substantiate that the internal jet formation correlates with the vortex ring generated after droplet coalescence. The essential mechanism for internal jet formation is further elucidated in Section VI, through the quantitative analysis of the vorticity generation at the gas–liquid interface, and the subsequent formation, growth, and detachment of the main vortex ring. In Section VII, a vortex-ring-based Reynolds number, characterizing the strength of the detached main vortex ring relative to the viscosity, is proposed as the criterion for the internal jet formation. Finally, the internal mixing is quantified in Section VIII by using a mixing index, and the resultant mixing rates are compared among different cases to demonstrate the crucial impact of the internal jet on the mixing performance of the merged droplet.

## II. NUMERICAL SPECIFICATIONS

We consider two initially stationary droplets, which are made to coalesce with each other at the point of contact at time $t = 0$, as illustrated in Figure 1(left). A cylindrical coordinate is established so that the line connecting the mass centers of the two spherical droplets forms the axial ($z-$) direction. The diameters of the smaller and the larger droplets are $D_S$ and $D_L$, respectively. The flows in both gas and liquid droplets are viscous and incompressible. The density and viscosity are $\rho_l$ and $\mu_l$ for the liquid, and $\rho_g$ and $\mu_g$ for the gas, respectively. The surface tension coefficient of the gas–liquid surface is $\sigma$. The tangential and normal directions in a local coordinate system established on the droplet interface are denoted by $\hat{t}$ and $\hat{n}$, respectively. The computational domain is an axisymmetric cylinder of $7D_S$ in length and $3D_S$ in radius. The outflow boundary conditions are specified on the all the boundaries except the axis. Our simulation results show that further extension of the domain has negligible influence on the results.

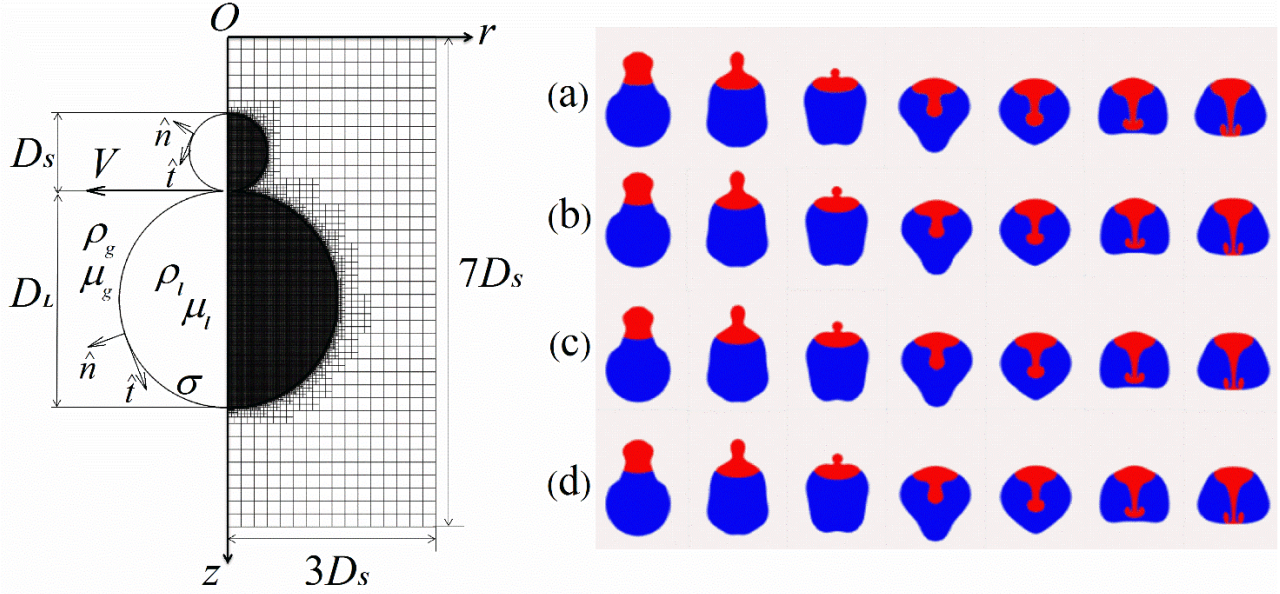

Fig.1 Computational domain and adaptive mesh for VOF simulation (left) and grid independence study (right). The symmetry boundary condition is specified on the symmetry axis 'z' and free outflow boundary conditions are specified on the other three boundaries. The adaptive mesh levels are (*a*) (6, 7, 8), (*b*) (6, 8, 8), (*c*) (7, 8, 9) and (*d*) (7, 9, 9), where the three numbers in parenthesis are the mesh levels for the gas zone, the fluid zone, and the interface zone, respectively.

Under the conditions that concern the present study, the gas–liquid density ratio $\rho_g/\rho_l$ is of $O(10^{-3})$ and gas–liquid viscosity ratio $\mu_g/\mu_l$ is of $O(10^{-2})$ so that they are assumed to have insignificant influence on the problem, as substantiated in the previous studies. Consequently, only two controlling non-dimensional parameters are considered, namely, the Ohnesorge number, which is defined as $Oh = \mu_l/\sqrt{\rho_l \sigma D_S}$ and measures the relative importance of viscous stress and surface tension, and the size ratio, defined as $\Delta = D_L/D_S$. In the present study, the Ohnesorge numbers are in the range of $8.3 \times 10^{-3} \sim 3.6 \times 10^{-2}$ and the size ratios varies from 1.0 to 3.2, which are similar to those considered in the experiments of Zhang *et al.*[43]. The characteristic length, time and velocity, $D_S$, $t_c = \sqrt{\rho_l D_S^3/\sigma}$ and $v_c = \sqrt{\sigma/\rho_l D_S}$, are used to nondimensionalize the present mathematical formulations. It is noted that $t_c$ is slightly larger than the natural oscillation time of the smaller droplet, $(\pi/4)\sqrt{\rho_l D_S^3/\sigma}$.

The volume of fluid (VOF) method adopted by the present study has been discussed in great detail[44, 45] and implemented in the open source code, Gerris[34, 35], which has been widely used in many multiphase flow problems[24, 46, 47]. The VOF method is characterized by the combination of the adaptive quad/octree spatial discretization, the geometrical VOF interface reconstruction, the continuum-surface-force surface tension formulation, and the height-function curvature estimation. In order to resolve the droplet interface and the internal flow within the droplet, the computational domain is divided into three zones, namely the gas, the interior of droplet and the interface. The mesh for each zone can be adaptively refined to a prescribed level, denoted by an integer $N$, at which the minimum cell size in the zone is of $O(2^{-N})$ of the zone dimension. Accordingly, we can use $(N_1, N_2, N_3)$ to describe the refinement levels in all the three zones. As an example, Figure 1 (left) shows an initial mesh at the refinement levels (5, 6, 7), in which the number of grid points is 13,851, equivalent to about 50,000 grid points in a uniform mesh system.

The grid independence of the computational results was examined in the study. Figures 1(a)-1(d) show the results for the coalescence of water droplets with $Oh = 8.28 \times 10^{-3}$ and $\Delta = 2.0$ by employing four different meshes. Figure 1(a) correspond to the mesh based on the refinement levels (6, 7, 8), which results in maximally 26,427 grid points, equivalent to about 200,000 on a uniform mesh. Figures 1(b)-1(d) correspond to the refinement levels (6, 8, 8), (7, 8, 9) and (7, 9, 9), respectively, which result in maximally 49,131, 79,215 and 173,451 grid points, respectively. It is seen that all the results are almost the same in terms of the evolution of droplet deformation, implying that the refinement level 6 for the gas zone and 8 for the interface zone provide adequate resolutions for the present problem. Figures 1(c)-1(d) show a finer structure of the vortex-ring-like mixing than Figure 1(b) after the non-dimensional time $T = t/t_c = 1.72$. This result suggests that the droplet interior should be sufficiently resolved to capture the features of the internal mixing. Based on the above results, the meshes generated at the refinement levels (6, 8, 8) were used in all the following computations. A typical simulation run up to $T = 4.0$ takes approximately five hours of real time on two Intel Xeon E5-2630 processors with 12 cores.

## III. EXPERIMENTAL VALIDATIONS

To validate the present numerical methods, the coalescence of two initially stationary droplets made of a mixture of silicon oil and bromobenzene in the environment of water was first simulated and compared with the experiment of Anilkumar *et al.*[30], as shown in Figure 2(a). The corresponding non-dimensional parameters are $Oh = 0.012, \Delta = 2.08, \rho_g/\rho_l = 10^{-3}$, and $\mu_g/\mu_l = 0.03$. It is seen that the simulation results are in good agreement with the experimental images in terms of the evolution of droplet deformation and that of the jet-like internal mixing. Anilkumar *et al.*'s experiment on the coalescence of two initially stationary droplets of silicone oil in water, corresponding to $Oh = 0.203, \Delta = 1.75, \rho_g/\rho_l = 1.27 \times 10^{-3}$, and $\mu_g/\mu_l = 10^{-3}$, shows that the jet-like mixing is absent because of the significantly increased droplet viscosity, as seen in Figure 2(b). The simulation results also reproduce the experimental observation in that the smaller droplet lodges on the larger one to form a dome shape.

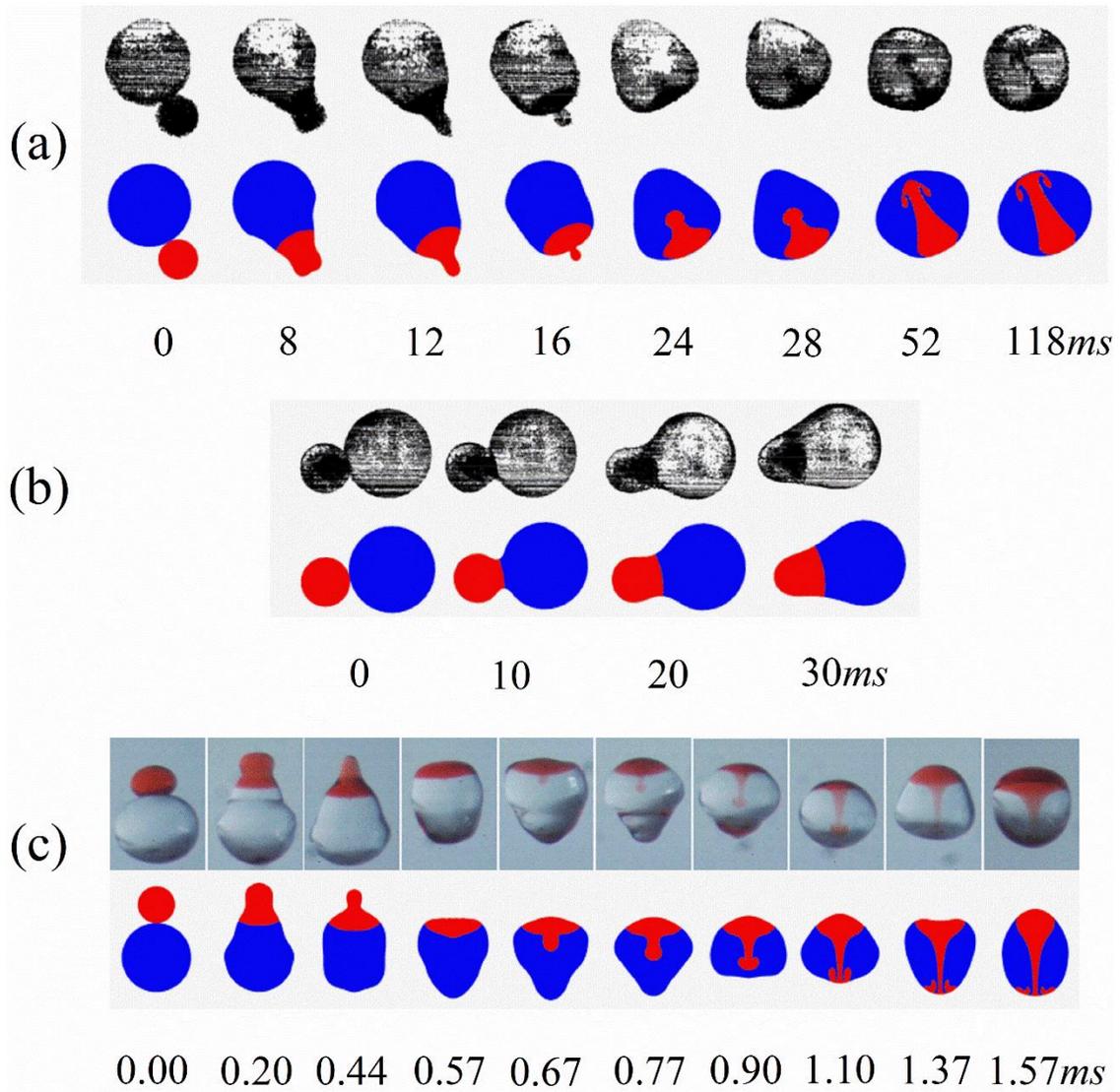

Fig.2 The internal mixing of two droplets of (a) a mixture of silicone oil and bromobenzene (3.3 CP measured viscosity), $We = 0.0, Oh = 0.012, \Delta = 2.08, t_c = 20.2ms$, (b) high viscous silicone oil (99.0CP measured viscosity), $We = 0.0, Oh = 0.203, \Delta = 1.75, t_c = 57.5ms$, and (c) water with $We = 0.47$ in experiment and $We = 0.0$ in simulation, $Oh = 8.29 \times 10^{-3}, \Delta = 1.89, t_c = 0.65ms$.

Figure 2(c) show the comparison between Tang *et al.*'s experiment[10] and the present simulation for the water droplet coalescence in atmospheric air with $Oh = 8.29 \times 10^{-3}, \Delta = 1.89, \rho_g/\rho_l = 1.2 \times 10^{-3}$, and $\mu_g/\mu_l = 1.8 \times 10^{-2}$. It is seen again that the simulation captures quantitatively well the droplet deformation and the internal mixing. It should be noted that the droplets coalesce at a non-zero albeit small Weber number ($We = 0.47$) in the experiment. Physically, a small duration of time is needed to drain out the gas film before coalescence occurs, but the droplet coalescence is triggered at $T = 0.0$ in the simulation. This can explain why the simulated droplets evolve slightly faster than the experimental ones. In summary, the present simulation methodology with the adopted mesh refinement is capable of capturing the detailed structure of internal mixing with quantitative agreement and hence will be used in the following simulation.

## IV. PHENOMENOLOGICAL DESCRIPTION OF INTERNAL JET-LIKE MIXING

### A. Representative case study

As a representative case, the coalescence of n-decane droplets was first investigated for a fixed Ohnesorge number of $Oh = 1.56 \times 10^{-2}$ and various size ratios. To show the long-time behavior of the internal mixing, the simulation lasts until $T = 4.0$. The small gas–liquid density ratio, $\rho_g/\rho_l = 1.68 \times 10^{-3}$, and the small gas–liquid viscosity ratio, $\mu_g/\mu_l = 1.96 \times 10^{-2}$, suggest that the gas inertia and viscosity are unlikely to cause any significant influence on the phenomena, as has been discussed in Section II. To facilitate our presentation, only three size ratios, namely, $\Delta = 1.8, 2.2$ and $2.8$ are shown in Figure 3(a). The mixing process is illustrated in the right half of each simulation image and the variation of local pressure, normalized by $3\sigma/D_S$, is shown in the left half image. Phenomenologically, the entire process from the commencement of droplet coalescence to $T = 4.0$ can be divided into four stages according to the characteristics of droplet deformation and jet formation.

Stage I (about $T = 0.00 \sim 0.20$): upon the droplet coalescence at the contact point, an interfacial ring cusp is formed in the vicinity of the point and tends to be smoothed out under the high capillary pressure. This causes a rapid, radially outward movement of the ring interface at the beginning stage of droplet coalescence. In the meantime, most of the mass in the droplets do not have noticeable movement along the axial direction. As a result, away from the vicinity of the ring cusp, the droplets remain almost spherical and the local pressures do not vary significantly.

Stage II (about $T = 0.20 \sim 0.40$): the mass in the smaller droplet is driven into the larger one under the capillary pressure difference of about $4\sigma(D_S^{-1} - D_L^{-1})$. A finger-like bulge with a round head is consequently formed on the far side of the smaller droplet, as clearly seen at $T = 0.40$. The radius of the round head of the bulge, denoted by $D_B$, is smaller than that of the initial droplet, and thereby results in larger capillary pressure (shown in red) in the bulge than that in the large droplet.

Stage III (about $T = 0.40 \sim 0.80$): the bulge merges into the droplet under the capillary pressure difference of about $4\sigma(D_B^{-1} - D_L^{-1})$. The mass of the bulge obtains an axial momentum in the vicinity of the axis as the result of the conversion of the surface energy of the bulge to its kinetic energy. After the bulge completely merges into the droplet, the pressure inside the droplet becomes almost uniform regardless of the slight local pressure variation due to the droplet oscillation.

Stage IV (about $T = 0.80 \sim 4.00$): the final stage is characterized by the long-time behavior of the jet-like mixing, which is affected by the viscous dissipation of the internal flow motion and the droplet oscillation. It is seen that all the cases show jet-like mixing patterns except the case of Δ = 1.8, in which the merged smaller droplet lodges on the larger one to form a dome shape during the entire stage.

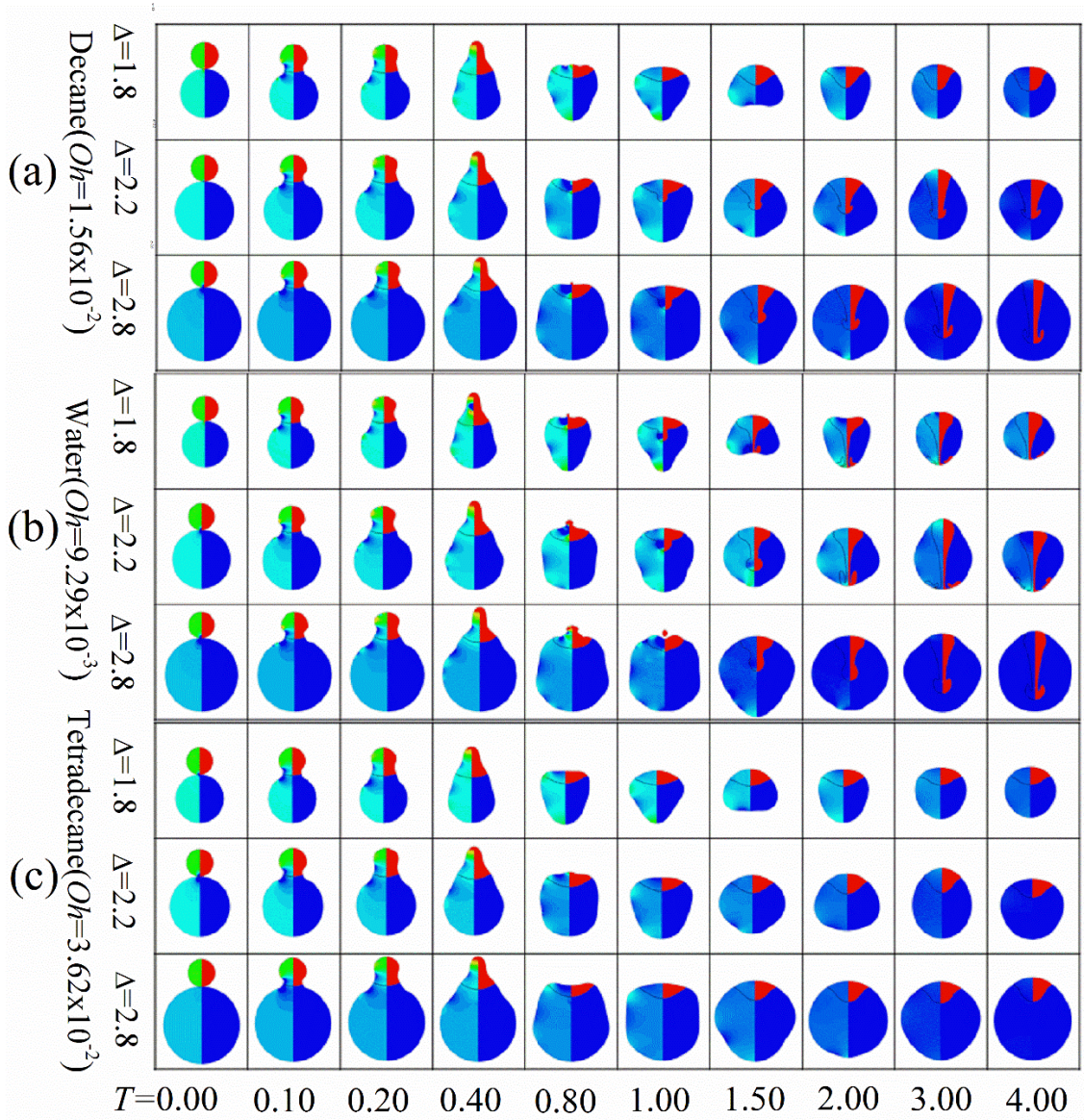

Fig.3 Deformation and mixing upon the coalescence of initially stationary droplets of (a) n-decane with $t_c = 0.50ms$, (b) water with $t_c = 0.33ms$, and (c) n-tetradecane with $t_c = 0.48ms$, and for different Δ ( Δ = 1.8, 2.2, 2.8). The left contour shows the static pressure, with the magnitude ranging from 0.0 of blue to 2.0 of red. The right contour plots the tracer variable, with red and blue tracking the fluid initially from the small and large droplets, respectively.

**B. Influences of $Oh$ and Δ**

For a comparative study of the phenomena, the above simulations were repeated for water and n-tetradecane, which provide experimentally realizable comparison with n-decane for studying the influences of surface tension and viscosity. Specifically, the viscosity of water ($8.9 \times 10^{-4} Pa \cdot s$) at room temperature is very close to that of n-decane ($8.5 \times 10^{-4} Pa \cdot s$), but the surface tension of water ($7.29 \times 10^{-2} N/m$) is almost three times larger than that of n-decane ($2.38 \times 10^{-2} N/m$). The surface tension of n-tetradecane ($2.65 \times 10^{-2} N/m$) is similar to that of n-decane, but its viscosity ($2.03 \times 10^{-3} Pa \cdot s$) is substantially larger than that of n-decane. The simulation results for the coalescence of water droplets at various size ratios and with $Oh = 8.29 \times 10^{-3}$, $\rho_g/\rho_l = 1.22 \times 10^{-3}$ and $\mu_g/\mu_l = 1.80 \times 10^{-2}$, are shown in Figure 3(b). The results for the coalescence of n-tetradecane droplets with $Oh = 3.62 \times 10^{-2}$, $\rho_g/\rho_l = 1.61 \times 10^{-3}$ and $\mu_g/\mu_l = 7.94 \times 10^{-3}$ are shown in Figure 3(c). Again, the influence of gas inertia and viscosity can be neglected because of the small gas–liquid density and viscosity ratios. It is noted that $Oh$ of n-tetradecane droplet almost doubles that of n-decane droplet, which in turn doubles that of water droplet. Several observations can be made by comparing the results shown in Figure 3, as follows.

First, owing to the smaller $Oh$ and thereby decreased viscous dissipation of water droplets, the mushroom-like jet emerges at $\Delta = 1.8$, as shown in Figure 3(*b*), but it is absent for either n-decane or n-tetradecane droplets at the same size ratio, as shown in Figure 3(a) and Figure 3(c), respectively. It is also noted that the jet-like mixing pattern is absent for n-tetradecane droplets at all the size ratios. These results accord with and extend the earlier experimental observation of Anikumar *et al.*[30] that the formation of "mushroom"-like jet is suppressed by increasing the droplet viscosity.

Second, for the cases with the emergence of the distinct jet-like mixing, water droplets display enhanced and faster mixing compared with n-decane droplets, as shown in Figure 3(b) compared with Figure 3(a) for $\Delta = 2.2$. It is seen that the jet penetrates the whole droplet and hits the bottom surface at $T = 2.0$ for the water droplet, but the jet penetration is about a half of the droplet size for the n-decane droplet at the same time. In view of the characteristic time $t_c$ of the water droplet is smaller than that of the n-decane droplet of the same size, the evolution of the jet-like mixing in the former is even faster and more substantial.

Third, an interesting albeit incidental phenomenon was observed for the water droplets of $\Delta = 2.8$, as shown in Figure 3(b). A thin neck forms on the bulge and pinches off the bulge to generate a satellite droplet, which subsequently re-merges into the "father" droplet, as shown at $T = 1.00$. Although the jet-like mixing pattern still emerges, it is not as prominent as the case without the pinch-off. In the experimental study of the pinch-off phenomenon during the coalescence of unequal-size droplets, Zhang *et al.*[43] found that there exists a critical $\Delta_{cr}$ for the emergence of pinch-off and that $\Delta_{cr}$ increases monotonically with $Oh$. For $Oh = 8.29 \times 10^{-3}$, the predicted $\Delta_{cr}$ is about 2.6, which slightly overshoots 2.3 observed in Zhang *et al.*'s experiments. The overshooting is caused by that the increasingly attenuated liquid neck is not sufficiently resolved in the present study. Physically, pinch-off occurs when two interfaces become sufficiently close to one another, being of $O(10)\ nm$, so that the van der Waals force triggers the interface collapse. Quantitatively predicting the pinch-off requires the numerical resolution of an extremely thin liquid neck, which is computationally challenging and therefore bypassed in the present study by limiting out scope to the size ratios below the critical ones determined by Zhang *et al.*'s experiment[43].

## C. Regime nomogram of jet-like mixing

The above comparative study has suggested that $Oh$ and $\Delta$ are crucial to the appearance of the internal mixing. In order to further quantify the influences of $Oh$ and $\Delta$ on the formation of the jet-like mixing pattern, we extended the simulations to wider ranges of these two parameters. The resulting $\Delta - Oh$ regime nomogram of the parametric study is shown in Figure 4, where the open triangles denote the cases with distinct jet-like mixing and the open squares those without jet-like mixing.

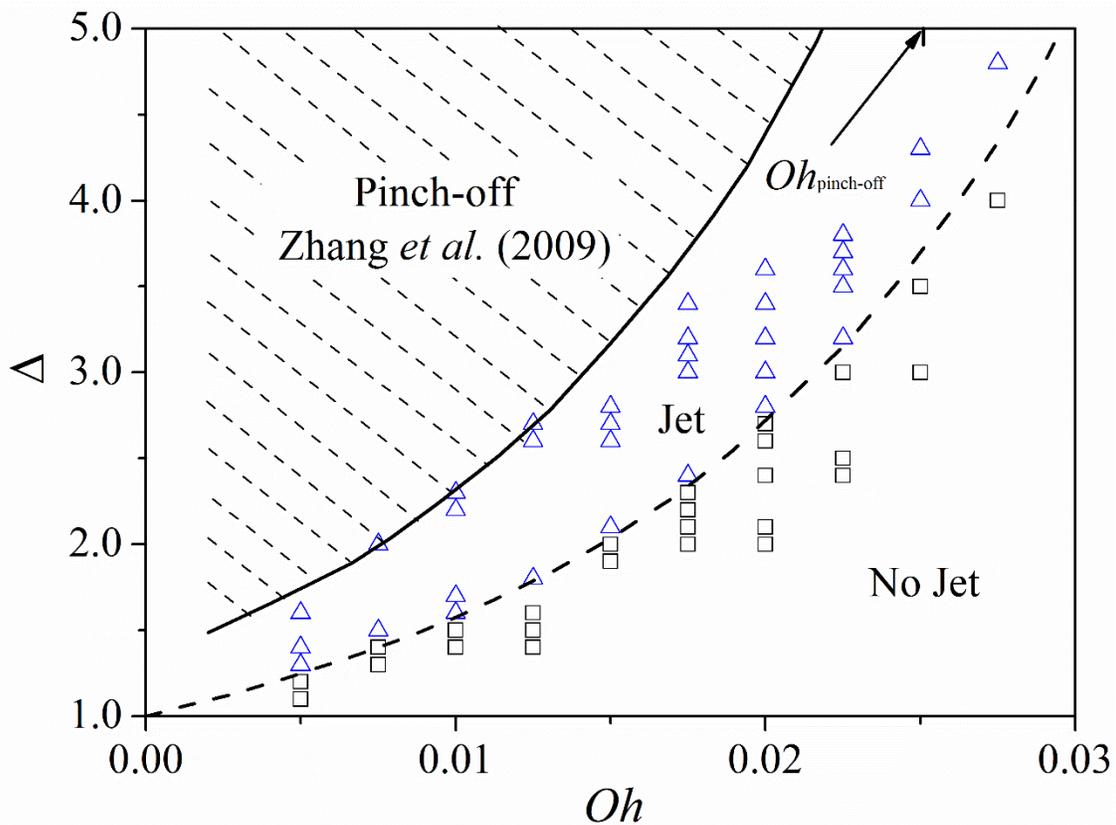

Fig.4 Regime nomogram of initially stationary droplet coalescence in the $\Delta - Oh$ parametric space. The jet-like mixing is denoted by the open triangle and otherwise by the open square. The dashed area corresponds to the ranges of $\Delta$ and $Oh$ effecting droplet pinch-off, which were measured experimentally by Zhang et al.[43]

For a given $Oh$, the jet-like mixing occurs as the size ratio is increased above a critical value indicated by the dashed line. Phenomenologically, the non-dimensional capillary pressure difference, $4(1 - \Delta^{-1})$, between the small and large droplets, increases with $\Delta$ and could be a driving force for the jet formation. Furthermore, Figure 3 also shows that as increasing the size ratio the coalescence-induced internal flow tends to be concentrated around the symmetry axis, which could serve as an additional source promoting the jet formation. It can be also observed that the critical Δ for the emergence of the internal jet increases with $Oh$, which again verifies the observations from Figure 3 that larger $Oh$ tends to suppress the jet formation.

For a complete phenomenological description of the problem, Zhang *et al.*'s experimental measurement on the emergence of droplet pinch-off is also shown in the figure as a shadow regime. It is noted that pinch-off does not occur for $Oh > 0.025$ in their experiment.

**V. IDENTIFICATION OF MAIN VORTEX RING SYNCHRONIZED WITH INTERNAL JET-LIKE MIXING**

**A. Characterization of main vortex ring**

We have demonstrated the dependence of the emergence of the internal jet on $Oh$ and Δ, however the basic mechanism of the jet formation still requires further investigation. As has been discussed in the Introduction, the generation of the jet-like mixing pattern should not be fundamentally different from other similar jetting phenomena, for example a starting vortex jet[48-50] formed due to the self-induced motion of a single vortex ring. This conjecture is partly substantiated in Figure 5 by the similar appearance of the jet-like structure and the vortex ring inside the merged droplet, characterized by normalized vorticity, $\omega^* = \omega t_c$. From previous studies[48-50], the formation of a starting jet can be boiled down to two essential problems: what is the source of vorticity forming the vortex ring? and how does the vortex ring detach from its source to form the jet? Similarly, we can illustrate the two important processes, namely, the generation of vorticity and the detachment of the main vortex ring, for the distinct internal jet subsequent to the droplet coalescence.

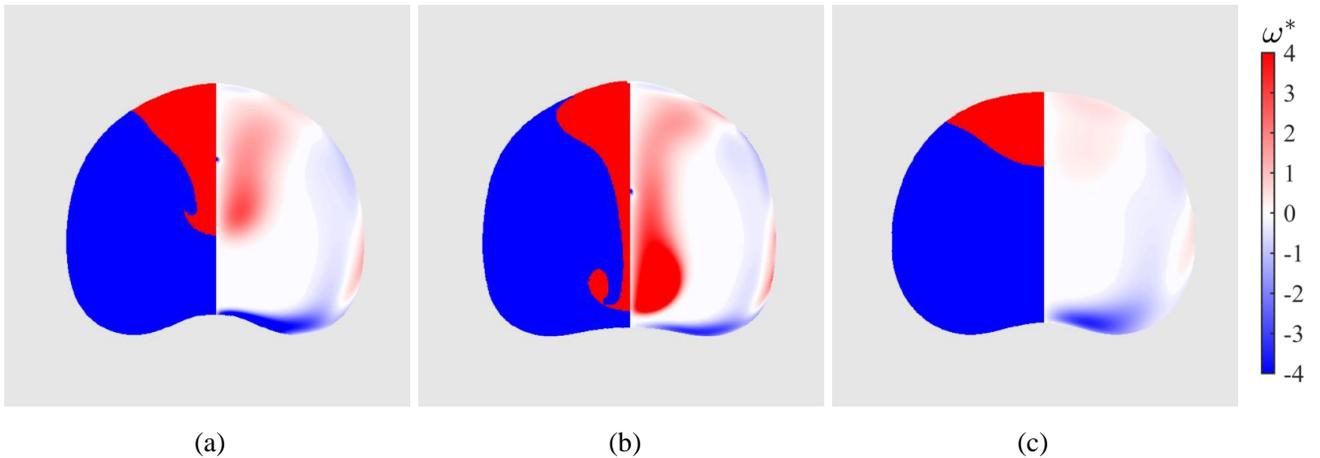

(a)          (b)          (c)

Fig.5 Comparison between the tracer variable distribution (left) and vorticity contour (right) at $T = 1.70$ for the cases of (a) n-decane, (b) water, and (c) n-tetradecane with $\Delta = 2.2$, shown in Figure 3. The color bar applies to the vorticity contours throughout the paper.

In order to study the generation and evolution of the main vortex ring that is responsible for the formation of the internal jet, one needs to first identify and track the region of the vortex. For the present problem having axisymmetry, we employ a simple vorticity-contour approach and consider the vorticity contour, $\omega = \omega_v$, encircling the main vortex as the vortex boundary, where $\omega_v$ is a small threshold vorticity. The results for a representative case (n-decane droplets with $\Delta = 2.2$) are presented in Figure 6, where the green contour on the left side of the droplet represents the identified boundary of the main vortex. By comparing the green contour with the vorticity contour on the right side of the droplet, we indeed verify the vorticity-contour approach to be effective in capturing the main vortex among other vortices emerging at different stages of droplet coalescence.

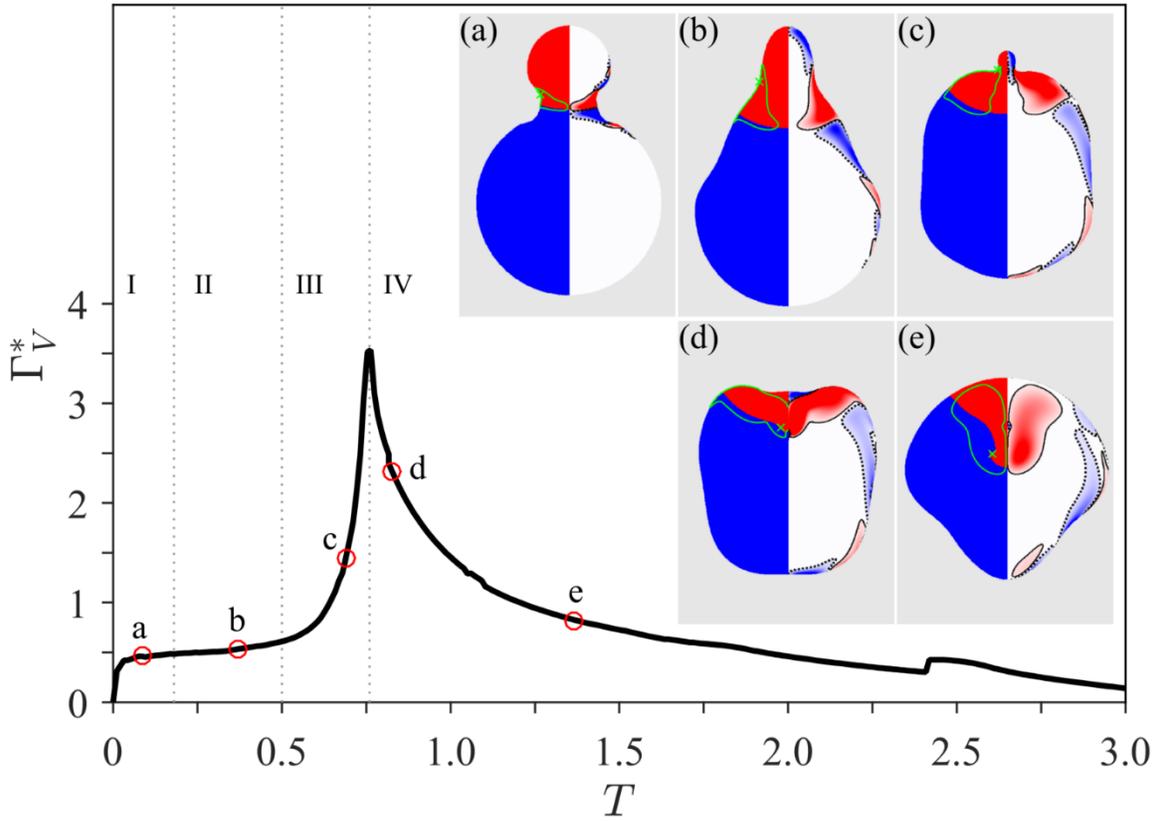

Fig.6 The main plot shows the evolution of the non-dimensional circulation of the main vortex for n-decane droplet coalescence with $\Delta = 2.2$, and $Oh = 1.56 \times 10^{-2}$. The subfigures (a)-(e) are the tracer variable plot (left) and vorticity contour (right) at the typical times (red circlets) on the vorticity evolution curve. In each subfigure, the green contour on the left side of the droplet represents the boundary of the main vortex identified through the vorticity-contour approach. The green cross marks the center of the main vortex corresponding to the local maximum vorticity. For clarity of illustration, positive and negative vortices in the vorticity contour are encircled by solid and dashed contour lines, respectively.

The identification of the main vortex region also enables the estimation of the total vorticity within the main vortex. The total vorticity is evaluated by calculating the circulation

$$\Gamma_V = \int_{A_V} \omega \, dA$$

where $A_V$ is the cross-section area of the main vortex. It should be noted that the volumetric integration over the volume, $V_V$, of the main vortex, $\int_{V_V} \omega \, dV$, is not an appropriate quantity for measuring the total vorticity. This is because Kelvin's circulation theorem[51, 52] dictates that $\Gamma_V$ is a conserved quantity in an ideal axisymmetric flow, but the volumetric integration is dependent on the radial distance of the vortex from the axis (Saffman[52], Eq. 1.5.22). It is seen in Figure 6 that $\Gamma_V$ first increases and then decreases at a slower rate, implying the existence of two physical processes: vorticity generation and vortex detachment, that will be discussed in detail in Section VI.

## B. Evolution of main vortex ring

Here, we shall discuss the evolution of the main vortex ring at different stages of droplet coalescence to further justify the correlation between the main vortex ring and the internal jet-like mixing. At Stage I when the initial droplet coalescence occurs, it can be observed from Figure 6(a) that vorticity develops only in the vicinity of the initial coalescence point, where surface deformation and its induced flow start. The rapid, outward expansion of the contacting region between the two droplets causes dramatical changes in the droplet geometry and velocity, thereby contributing to a fast rate of vorticity generation.

At Stage II when the initial coalescence region has sufficiently expanded as shown in Figure 6(b), the initial conversion from surface energy to kinetic energy gradually reaches stagnant. As a result, the velocity and the geometry of the droplet enter a slow-variation stage, causing a gradual vorticity generation inside the main vortex. By the end of this stage, the remainder of the smaller droplet becomes an axial finger-like liquid bulge, forming a new neck which would lead to the next peak of surface energy discharge.

At Stage III, the fluid bulge collapses into the large droplet under surface tension as shown in Figure 6(c). Similar to Stage I, a substantial amount of surface energy of the merged interface is rapidly converted to the kinetic energy of the induced flow around the confined neck region of the small fluid bulge, resulting in massive vorticity production in the main vortex.

At Stage IV, after the complete merge of the fluid bulge into the large droplet, the main vortex ring starts to detach from the droplet surface and eventually forms the internal jet, as shown in Figure 6(d) and Figure 6(e). During this detaching process, the circulation of the main vortex ring first declines rapidly because of the diffusion of negative vorticity from the surface, and then drops gradually because of viscous dissipation after the complete detachment of the vortex ring.

## VI. FORMATION, GROWTH AND DETACHMENT OF MAIN VORTEX RING

### A. Formation of main vortex ring in stage I

We have demonstrated the dependence of the emergence of the internal jet on $Oh$ and $\Delta$, however the basic mechanism of the jet formation still requires further investigation. As has been discussed in the Introduction, the generation of the jet-like mixing pattern should not be fundamentally different from other similar jetting phenomena, for example a starting vortex jet[48-50] formed due to the self-induced motion of a single vortex ring. This conjecture is partly substantiated in Figure 5 by the similar appearance of the jet-like structure and the vortex ring inside the merged droplet, characterized by normalized vorticity, $\omega^* = \omega t_c$. From previous studies[48-50], the formation of a starting jet can be boiled down to two essential problems: what is the source of vorticity forming the vortex ring? and how does the vortex ring detach from its source to form the jet? Similarly, we can illustrate the two important processes, namely, the generation of vorticity and the detachment of the main vortex ring, for the distinct internal jet subsequent to the droplet coalescence.

We first focus on the process of vorticity generation associated with the main vortex. According to Helmholtz's third theorem[52] or Kelvin's circulation theorem[52, 53], vorticity cannot be created inside of a fluid, except at the boundaries or interfaces. Therefore, the initial generation of vorticity associated with the main vortex must happen at the gas–liquid interface. For the present problem, as has been discussed in the Introduction, the significantly smaller density and viscosity of gas compared with those of liquid enable us to approximate the gas-liquid interface as a free surface with a vanishing shear stress[40, 48, 49] on it. In the reference frame defined in Figure 6, the shear stress has the form

$$\sigma = \hat{\boldsymbol{n}} \cdot \nabla \boldsymbol{u} \cdot \hat{\boldsymbol{t}} + \hat{\boldsymbol{t}} \cdot \nabla \boldsymbol{u} \cdot \hat{\boldsymbol{n}} = 0. \quad (1)$$

and hence

$$\omega = \hat{\boldsymbol{n}} \cdot \nabla \boldsymbol{u} \cdot \hat{\boldsymbol{t}} - \hat{\boldsymbol{t}} \cdot \nabla \boldsymbol{u} \cdot \hat{\boldsymbol{n}} = -2\hat{\boldsymbol{t}} \cdot \nabla \boldsymbol{u} \cdot \hat{\boldsymbol{n}} \neq 0. \quad (2)$$

According to Lundgren and Koumoutsakos[41], the interface vorticity in equation (2) could be further decomposed into two parts by

$$\omega = -2\frac{\partial \boldsymbol{u}}{\partial s} \cdot \hat{\boldsymbol{n}} = -2\frac{\partial u_n}{\partial s} + 2\kappa u_t, \quad (3)$$

where $\kappa = \hat{\boldsymbol{t}} \cdot \partial \hat{\boldsymbol{n}}/\partial s$ is the local curvature of the surface; $u_n$ and $u_t$ are the velocity components normal and tangential to the surface, respectively. Equation (3) suggests that non-zero vorticity would be generated from interfacial flow induced by surface deformation. For steady flows where the free surface is stationary, $u_n$ becomes zero and the first term of Equation (3) could be dropped out. For the present problem that is highly unsteady and involves significant deformation of the interface, both terms of equation (3) individually affects the vorticity generation of the main vortex at different stages.

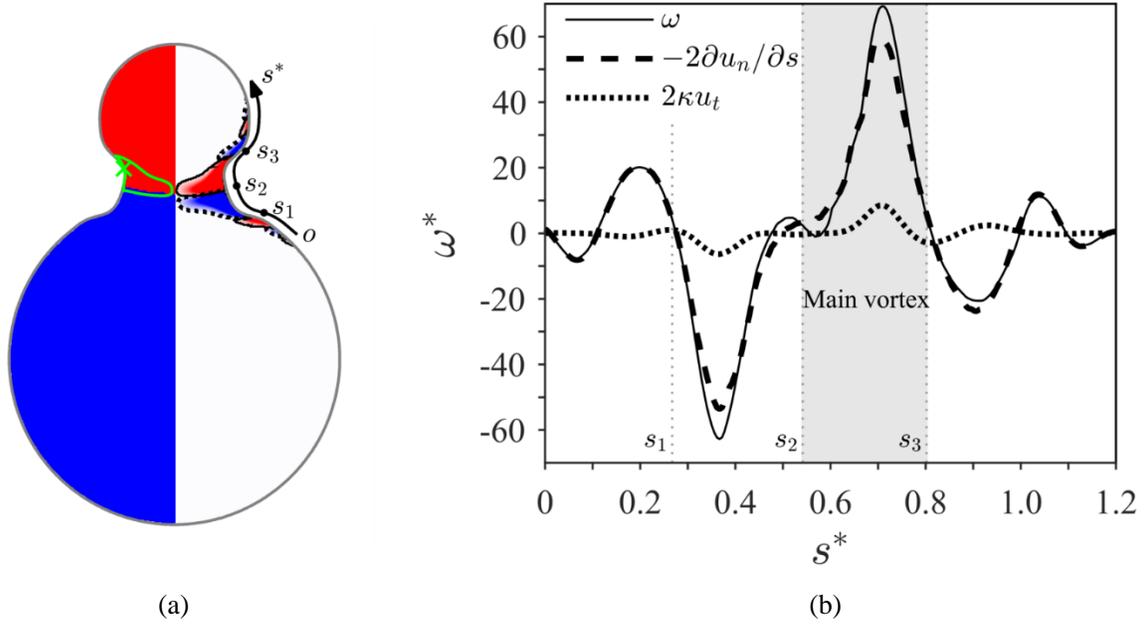

(a)            (b)

Fig.7 (a) The tracer variable plot (left) and vorticity contour (right) at a representative instant of Stage I (Fig. 6a) for n-decane droplet coalescence with $\Delta = 2.2$, and $Oh = 1.56 \times 10^{-2}$. The green contour marks the boundary of the main vortex. (b) The vorticity distribution along the gas–liquid interface near the main vortex. The $s^*$ coordinate is along the surface of the droplet, as indicated by the black arrow in (a).

    Figure 7 shows the distribution of vorticity at the gas–liquid interface where massive vorticity is being generated initially inside the main vortex at Stage I. We can observe that a main peak of positive vorticity is located between $s_2$ and $s_3$, which corresponds to the portion of the interface overlapping with the boundary of the main vortex. This verifies that the growth of the total vorticity inside the main vortex shown in Figure 6 is indeed related to the non-zero vorticity generation at the interface. Furthermore, the vorticity contributions from the two terms of Equation (3) are also plotted in Figure 7. It is seen that the first term, $-2\partial u_n/\partial s$, almost equals the vorticity itself, and the second term, $2\kappa u_t$, is negligible. This implies that the initial vortex generation in Stage I is mainly caused by the normal velocity gradient in the tangential direction of the outwardly moving surface, as the movement of surface is normal to the surface and $u_n$ decreases from $s_2$ to $s_3$. It is interesting to note that vorticity generation is approximately symmetric with respect to the contacting plane of the two droplets during Stage I. Specifically, a counter-rotating vortex is created on the large droplet side from the surface between $s_1$ and $s_2$, with the $-2\partial u_n/\partial s$ term again dominating the vorticity generation. Although this "negative" vortex is only slightly weaker than the main vortex, it gradually decays in later stages, whereas the main vortex will be further enhanced.

**B. Growth of main vortex ring in stage III**

In Section V, we have seen that Stage III corresponds to the primary vorticity generation phase for the main vortex. To understand the underlying mechanism of vortex growth during this stage, Figure 8 plots the vorticity distribution along the gas–liquid interface at the instant corresponding to Figure 6(c). Here, the curve between $s_1$ and $s_2$ represents the portion of the gas–fluid interface associated with the main vortex. It can be observed that major vorticity generation occurs in the vicinity of the connecting point between the bulge and the main droplet. The two vorticity components in Equation (3) are also plotted in Figure 8, and their peaks are comparable to each other along the interface of the main vortex. This indicates that both the interface deformation and the large curvature equally contribute to the significant vorticity generation during this stage.

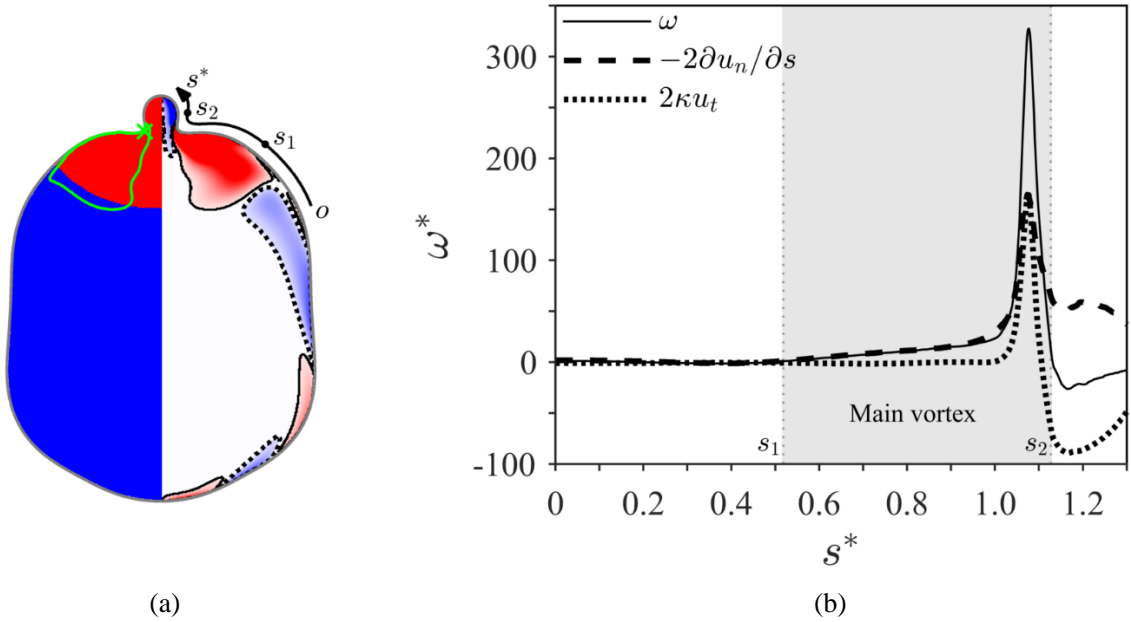

(a)          (b)

Fig.8 (a) The tracer variable plot (left) and vorticity contour (right) for a representative instant of Stage III (Fig. 6c) for n-decane droplet coalescence with $\Delta = 2.2$ and $Oh = 1.56 \times 10^{-2}$. (b) The corresponding vorticity distribution along the gas–liquid interface near the main vortex.

It is also noted that, in the region beyond the main vortex and towards the axis, the tangential velocity and curvature combined generate negative vorticity, which exceeds the positive vorticity generated by interface deformation, causing the emergence of a negative vortex inside the small bulge.

**C. Detachment of main vortex ring in stage IV**

The above analysis answers the question about where the vorticity inside the internal jet comes from, but it does not explain how the concentrated vorticity becomes a self-induced vortex ring that subsequently translates within the droplet to form an internal jet. In other words, the cumulation of vorticity inside a vortex ring does not guarantee the formation of a jet. Taking the starting jet problem as an example, the jet formation criterion requires that the non-dimensional formation time is larger than a certain value[49] (a typical number is 4), when the vortex ring is strong enough to "rip-off" its own vorticity-feeding shear layer; otherwise, the vorticity would keep growing inside the vortex ring without detachment and thereby no jet can be formed.

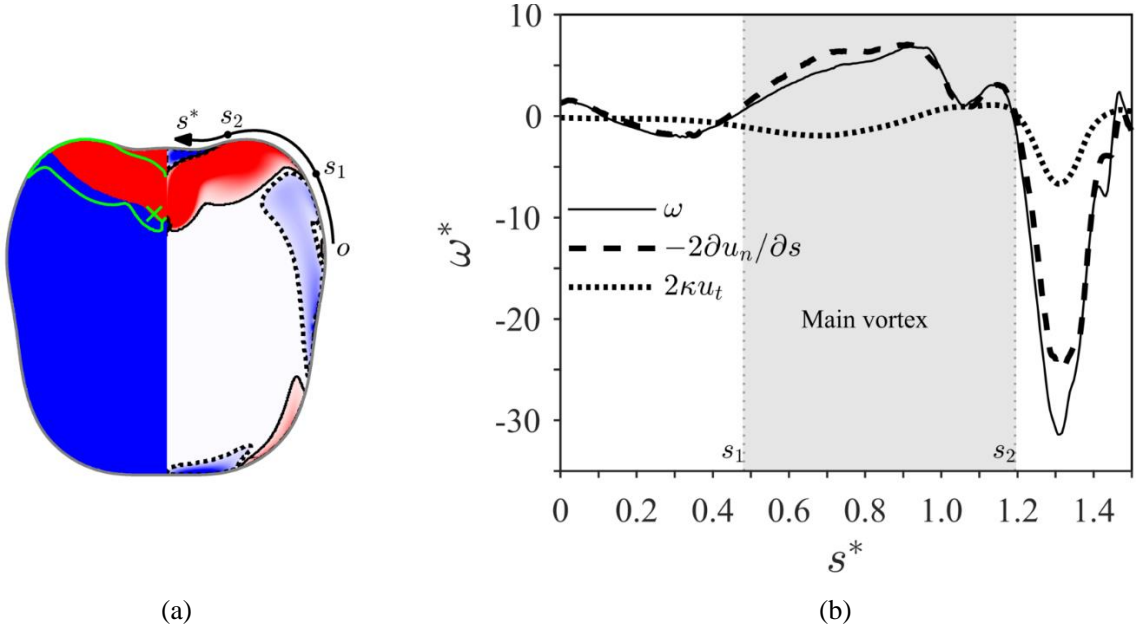

(a) (b)

Fig.9 (a) The tracer variable plot (left) and vorticity contour (right) for a representative instant of Stage IV (Fig. 6d) for n-decane droplet coalescence with $\Delta = 2.2$ and $Oh = 1.56 \times 10^{-2}$. (b) The corresponding vorticity distribution along the gas–liquid interface near the main vortex.

For the formation of the internal jet due to droplet coalescence, we hypothesize that a similar mechanism exists such that the internal jet does not form until the main vortex ring detaches from the surface of the droplet. Consequently, there must be an additional mechanism that "peels" the original vortex ring off the surface. This problem resembles that of flow separation on a smooth body, where vorticity inside the boundary layer detaches from the surface to become a free vortex in the wake. It is well-known that flow separation happens where the adverse pressure gradient is strong enough to generate vorticity opposite to the original shear layer. Analogically, we hypothesize that the emergence of negative vorticity at the gas-liquid interface would be responsible for the detachment of the main vortex ring.

To verify this hypothesis, we plot the vorticity distribution along the surface of the droplet at the beginning of Stage IV, as shown in Figure 9. It is confirmed that the main vortex starts to detach from the central portion of the interface ($s^* > s_2$) where significant negative vorticity is generated. Moreover, Figure 9 also shows that the negative vorticity generation near the axis is largely attributed to the first term of Equation (3), $-2\partial u_n/\partial s$. We further recognize that this negative vorticity generation actually corresponds to an outward retraction of the dented surface subsequent to the complete merge of the droplet. In analogy to flow separation, $s_2$ can be considered as a dynamic flow separation point, which moves in the radial direction as the negative vortex region expands out, and cut off the main vortex from the surface in a razor-like fashion. After the main vortex completely detaches from the surface, it becomes a free vortex ring translating with its self-induced velocity to form an internal jet.

**VII. VORTEX-RING-CRITERION FOR INTERNAL JET FORMATION**

Previous sections have demonstrated that the emergence of the internal jet after droplet coalescence is closely related to the generation and detachment of the main vortex. In this section, we shall seek quantitative criterion of the internal jet based on the main vortex ring.

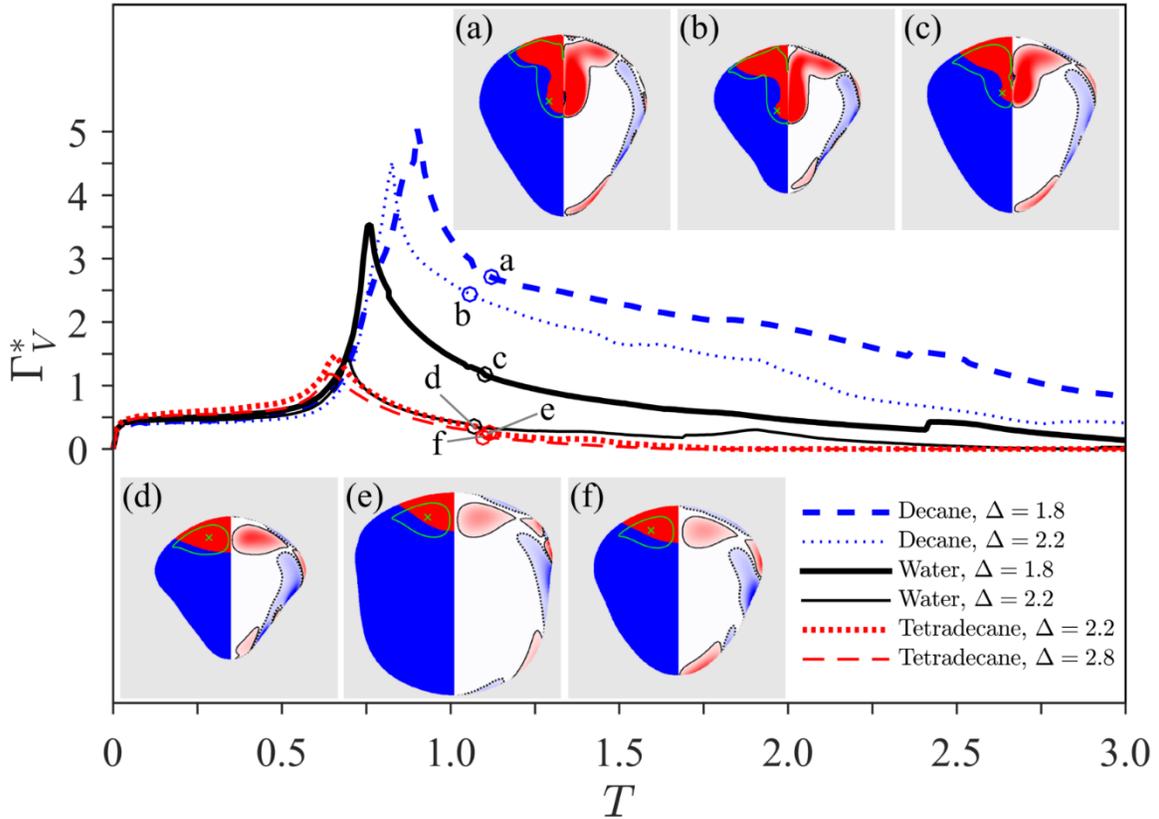

Fig.10 Evolution of the non-dimensional circulation around the main vortex for different cases of droplet coalescence. The subfigures (a)-(f) present the tracer variable plot (left) and vorticity contour (right) for the instants when the main vortex completely detaches from the surface of the droplet. The circulations corresponding to these instants are also marked on the evolution curves.

We start by comparing the evolution of the main vortex between cases with the internal jet (water droplets with $\Delta = 1.8, 2.2$ and n-decane droplets with $\Delta = 2.2$) and those without the internal jet (n-decane droplets with $\Delta = 1.8$ and n-tetradecane droplets with $\Delta = 1.8, 2.2$), as shown in Figure 10. The droplet configurations and vorticity contours at the instant when the vortex ring completely detaches from the droplet surface are also presented in the figure, with their corresponding total vorticities being marked on the circulation evolution curves, respectively. It is seen that, at the instant of vortex detachment, the cases without jet formation have significantly lower total vorticity compared with the cases with jet formation. Moreover, the lower circulation for the cases without jet formation should be attributed to either shorter period or smaller rate of vorticity generation during Stage III. It is noted that the less total vorticity generated during Stage III is related to the diminished bulge structure created at the end of Stage II, which does not possess enough surface energy to be converted during Stage III.

The above analysis qualitatively indicates that the internal jet appears only if the vortex ring is strong enough by the end of the vortex detachment. Physically, an ideal free vortex ring would readily translate downstream and form a jet under its self-induced flow, unless the viscosity significantly dissipates the vortex ring. Consequently, the formation of the internal jet after the vortex detachment is determined by the competition between the advective momentum associated with the vortex ring and the momentum lost due to viscous diffusion. To characterize this competition, a Reynolds number can be defined by

$$Re_J = \frac{U_V r_V}{\nu}, \tag{4}$$

where $U_V$ and $r_V$ are the characteristic translational velocity and the radius of the free vortex ring, respectively. Based on the dynamics of a vortex ring[52], $U_V$ can be expressed as

$$U_V = \frac{\Gamma_V}{4\pi r_V}. \tag{5}$$

Combining Equations (4) and (5) gives

$$Re_J = \frac{\Gamma_{VD}}{4\pi \nu} = \frac{\Gamma_{VD}^*}{4\pi Oh}, \tag{6}$$

where $\Gamma_{VD}$ (and its non-dimensional form $\Gamma_{VD}^*$) represents the circulation of the main vortex ring when it completely detaches from the droplet surface.

With $\Gamma_{VD}^*$ calculated in a similar approach to Figure 10 (circulations associated with the circlets), Equation (6) can be validated for different cases of droplet coalescence, as shown in the $Oh - \Gamma_{VD}^*$ regime nomogram in Figure 11. The mushroom-like jet formation occurs with about $Re_J > 4.12$, no jet formation with $Re_J < 2.41$, and in-between is a transition regime. The transition from the "no-jet" regime to the "mushroom-like jet" regime with increasing $Re_J$ is manifested by the five typical cases from 1 to 5, as shown in the subfigures of Figure 11. Thus, the jet number $Re_J$ serves as a phenomenological criterion for jet formation subsequent to the coalescence of two initially stationary droplets.

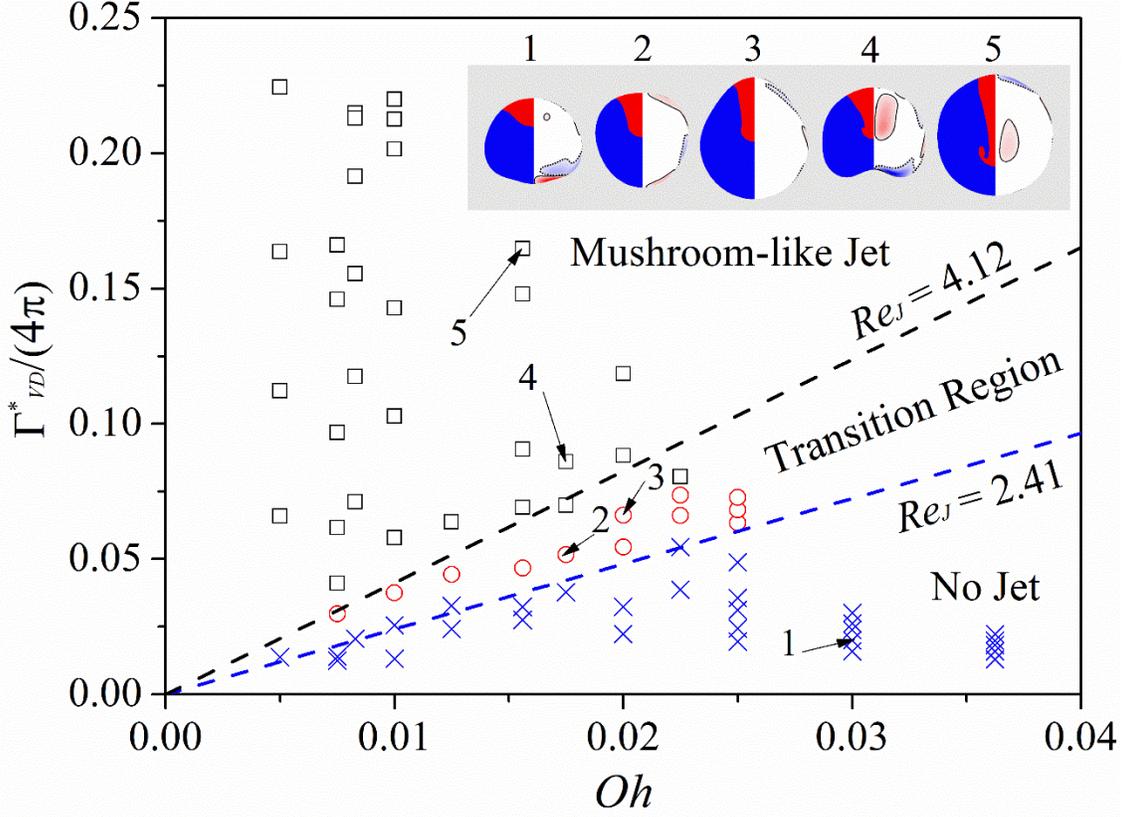

Fig.11 Validation of the vortex criterion for the jet formation for different cases of droplet coalescence. The slopes of the two dashed lines correspond to $Re_J = 2.41$ and $Re_J = 4.12$, marking the transition from the regime without jet formation to that with mushroom-like jet formation.

**VIII. CORRELATION BETWEEN VORTEX-INDUCED JET AND INTERNAL MIXING**

As discussed in the Introduction, the present work was motivated to understand the influence of the jet-like structure on internal mixing. The comparative study in Section 4 has qualitatively demonstrated that mixing between the two droplets is intimately related to the mushroom-like jet. Specifically, the occurrence of the jet seems to cause large stretching and roll-up of the interfaces of two droplets, thereby increasing the contacting surface and contributing to enhanced mixing performance. To further quantify this observation on the internal mixing within the merged droplet, a mixing index $M$ independent of $Oh$ and $\Delta$ can be defined as follows[10, 54],

$$M = 1 - \frac{\int_V |C - C_\infty| H(f-1) dV}{\int_V |C_0 - C_\infty| H(f-1) dV}, \tag{7}$$

where the VOF function $f = 0$ in the gas and $f = 1$ in the droplets; the Heaviside step function $H(f-1)$ limits the integration domain to be within the droplets. $C$, $C_0$, and $C_\infty$ are the distribution functions of the time-dependent "concentration" of the small droplet liquid in the merged mass, in the unmixed droplets, and in the fully mixed mass, respectively, and are defined by

$$C = |2\phi - 1|C_0, \quad C_0 = \begin{cases} 0, \phi = 0 \\ 1, \phi > 0 \end{cases}, \quad C_\infty = 1/(1 + \Delta^3), \tag{8}$$

where $\phi$ is the spatially and temporally varying dye function, which is defined as 1 in the small droplet and 0 in the large droplet. Therefore, $M$ varies between 0 and 1 with larger value indicating better mixing.

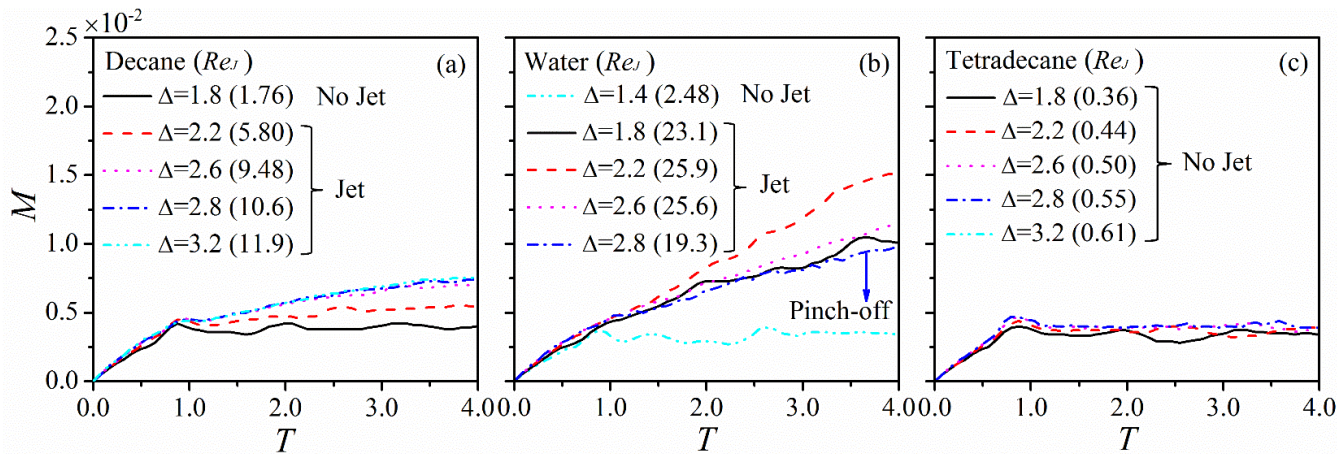

Fig.12 Time evolution of the mixing index for n-decane ($Oh = 1.56 \times 10^{-2}$), water ($Oh = 8.29 \times 10^{-3}$), and n-tetradecane ($Oh = 3.62 \times 10^{-2}$) droplets of various size ratios. $Re_J$ for each case is presented in the parentheses to relate the performance of mixing to the formation of internal jet.

The time evolution of the calculated $M$ for n-decane, water and n-tetradecane and for various $\Delta$ are shown in Figure 12. The slope of the curves quantifies the rate of mixing at different times. Several observations can be made about the mixing processes.

First, for all the cases, the mixing rate seems to be independent of $Oh$ or $\Delta$ at Stages I-III (up to around $T = 0.8$), when the small droplet has not completely merged into the larger one. This is because mixing in the early stages are mostly caused by the geometrical changes of the two-droplet configuration. These changes are associated with the coalescence of interfaces around the initial contacting point (Stage I), the expansion of the contacting surface (Stage II), and the collapse of the liquid bulge (Stage III), which to a considerable extent are similar processes among cases with different $Oh$ or $\Delta$.

Second, the increase of $M$ during Stage IV substantially varies with $Oh$ and $\Delta$, suggesting the important role of the jet-like structure that does not present until this stage. The values of $Re_J$ for different cases are presented in the parentheses to indicate the correlation between jet formation and internal mixing. It is seen that, for small $Re_J$ cases where no internal jet is formed, the average mixing rate is approximately zero in Stage IV and is almost independent of $Re_J$. Together with the above-discussed mixing independency during Stages I-III, it seems that mixing is intrinsically self-similar throughout the merging process if no internal jet presents. This, in turn, provides a direct evidence that the vortex-induced jet has a significant impact on the internal mixing of the merged droplet.

Third, for those cases where mushroom-like jet forms, $Re_J$ measures the relative strength of the vortex ring associated with the internal jet. Therefore, larger $Re_J$ corresponds to a stronger jet, and in turn indicates larger convective momentum or lower viscous dissipation or both; enhanced mixing rate is consequently expected. This can be readily verified by the n-decane cases shown in Figure 12(a), where the average mixing rate seems to monotonically increase with $Re_J$. Furthermore, comparing the n-decane cases with the water cases shown in Figure 12(b), we can also confirm that the water cases in general has larger mixing rates than the n-decane cases, as the water cases have larger $Re_J$.

Forth, it is noted that the monotonic dependence of the mixing rate on $Re_J$ seems to not hold exactly for the water cases. Specifically, the $\Delta = 2.2$ case has much larger mixing rate than the $\Delta = 2.6$ case, although $Re_J$ for the two cases are nearly the same. This can be attributed to the tendency of droplet pinch-off for the water case with $\Delta = 2.6$, where the surface energy responsible for the interface oscillations at the later stage of internal mixing is significantly reduced. Moreover, for the cases with droplet pinch-off, the long-term mixing performance is also affected by the surface oscillations after the jet has penetrated through the merged droplet and hit the gas–liquid interface on the opposite side.

Finally, for the cases with internal jet but without droplet pinch-off, the average mixing rate also monotonically increases with $\Delta$ or decreases with $Oh$, which is consistent with our observation from Figure 3 that the jet-like mixing becomes more prominent with increasing $\Delta$ or decreasing $Oh$.

## IX. CONCLUDING REMARKS

In this paper, the coalescence of two initially stationary droplets of unequal sizes was investigated numerically using an improved VOF method and adaptive mesh refinement algorithm, with emphasis on two key problems: the formation of the internal jet and its effect on mixing between the two droplets.

Phenomenologically, the jet-like mixing was perceived based on three representative fluids, water, n-decane, and n-tetradecane. It was found that both decreasing the fluid viscosity and increasing the size ratio would facilitate the emergence of the jet-like mixing. This observation was further justified by a regime nomogram describing the occurrence of the jet-like mixing with varying size ratio and Ohnesorge number.

The fundamental mechanism responsible for forming the internal jet was understood by first identifying the correlation between the jet-like structure and the main vortex ring inside the merged droplet. Based on the evolution of the main vortex ring, we identified three stages crucial to the jet formation, namely, vortex-ring formation, vortex-ring growth, and vortex-ring detachment. The initial vortex generation upon droplet coalescence, albeit being small in magnitude, is mainly caused by the normal velocity gradient in the tangential direction of the outwardly merging surface around the coalescence point. The rapid collapse of the small bulge into the large droplet, which results in dramatical interface deformation and local large-curvature interface, causes massive vorticity generation and therefore vortex-ring growth. The outward retraction of the dented surface promotes generation of negative vorticity that detaches the main vortex ring from the surface to form a self-induced jet.

Furthermore, an analysis concerning the circulation of the main vortex ring was performed by comparing cases with and without the internal jet. The result points to a positive correlation between the occurrence of the internal jet and the circulation of the main vortex ring after its complete detachment. Recognizing that the formation of the internal jet is the outcome of a competition between the convective momentum carried with the vortex ring and the viscous dissipation, we defined a special Reynolds number, $Re_J$, as the criterion for jet formation. It was verified through various droplet coalescence cases that $Re_J > 4.12$ approximately gives the mushroom-like jet and $Re_J < 2.41$ predicts no jet formation, with in-between marking the transition regime.

A mixing index was defined to quantify the effect of the internal jet on mixing. For the cases without internal jet formation, the average mixing rate has minor dependence on $\Delta$ or $Oh$, reflecting an intrinsic self-similarity associated with the configuration evolution of the two-droplet system during their coalescence. For the cases with internal jet formation, the average mixing rates are however significantly enhanced and the mixing index tends to monotonically increase with increasing $\Delta$ or decreasing $Oh$.

Because $Re_J$ physically characterizes the relative strength of the detached vortex ring, which controls the strength of the forming jet, we compared $Re_J$ among cases of different $\Delta$s or $Oh$s to seek a direct correlation between the vortex-induced jet and internal mixing. It was confirmed that the average mixing rate of the vortex-ring detachment stage grows monotonically with $Re_J$ for large $Re_J$ cases with jet formation. It should be noted that the above observations and conclusions only hold for the cases without droplet pinch-off or the tendency of pinch-off, because pinch-off will dramatically reduce the surface energy driving the interface oscillation, which affects the mixing in the stage of jet formation.

**ACKNOWLEDGMENTS**

This work was supported by the Hong Kong RGC/GRF (operating under contract numbers PolyU 152217/14E and 152651/16E) and partly by the Hong Kong Polytechnic University (G-UA2M and G-YBGA).